# Barycentric Coordinates as Interpolants

Russell A. Brown

**Abstract**

Barycentric coordinates are frequently used as interpolants to shade computer graphics images. A simple equation transforms barycentric coordinates from screen space into eye space in order to undo the perspective transformation and permit accurate interpolative shading of texture maps. This technique is amenable to computation using a block-normalized integer representation.

**Mathematical Formulation**

Consider the triangle $P_0 P_1 P_2$

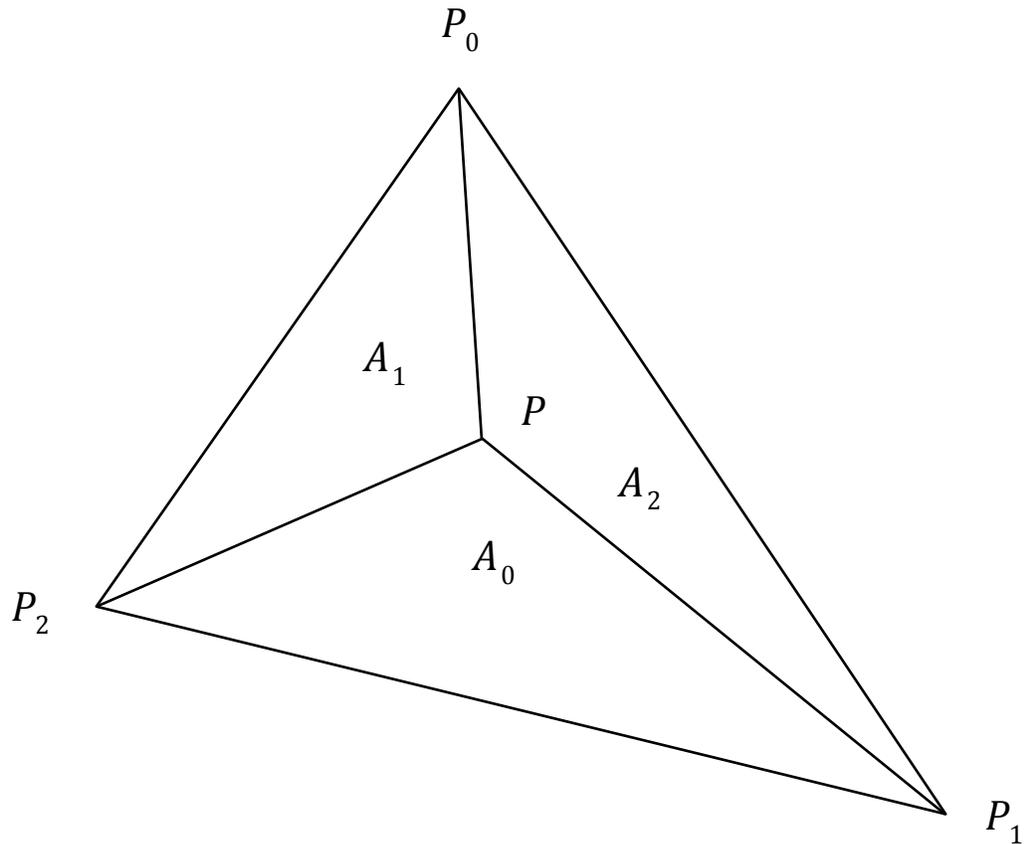

The points $P_0$, $P_1$ and $P_2$ are the vertices of the triangle. $P$ is a point at which interpolation will be performed. Assume that point $P_0$ has Cartesian coordinates $(x_0, y_0, z_0)$, color coordinates $(r_0, g_0, b_0)$ and texture coordinates $(u_0, v_0)$. Points $P_1$ and $P_2$ are similarly defined. The Cartesian coordinates are defined in perspective space, *i.e.*, they have been multiplied by a perspective matrix and perspective divided. The color and texture coordinates are defined in world space. We use the $w$-coordinate instead of $z$. This convention is used because $w$ represents eye space $z$, *i.e.*, the $z$-coordinate by which eye space $(X, Y)$ coordinates are divided to produce perspective space $(x, y)$ coordinates. As will be seen shortly, $w$ may be used to undo the perspective division so that interpolation occurs in eye space, which is equivalent to interpolation in world space. Moreover, for depth priority calculations, $w$ may be preferred to perspective $z$ when range priority calculations use a fixed-point representation [1].

The first step of the shading calculation is to construct a set of interpolants for point $P$ situated at a particular pixel. Barycentric coordinate [2] interpolants may be calculated in a simple manner from the points $P_0$, $P_1$ and $P_2$ by defining

$A_0$ to be the screen-space area of triangle $PP_1P_2$
$A_1$ to be the screen-space area of triangle $PP_2P_0$
$A_2$ to be the screen-space area of triangle $PP_0P_1$

which permits calculation of the barycentric coordinates $b_0$, $b_1$ and $b_2$ as

$$b_0 = A_0/(A_0 + A_1 + A_2)$$
$$b_1 = A_1/(A_0 + A_1 + A_2) \qquad (1)$$
$$b_2 = A_2/(A_0 + A_1 + A_2)$$

Because division is performed using the sum of triangle areas, the sum of the barycentric coordinates $b_0 + b_1 + b_2 = 1$. Also, if any $b_i < 0$, the pixel is outside the edge opposite $P_i$, a fact which can be useful in scan conversion.

The areas of triangles $PP_1P_2$, $PP_2P_0$ and $PP_0P_1$ may be computed in a trivial manner. For example, the area $A_0$ of triangle $PP_1P_2$ is equal to one-half the magnitude [3] of the cross product $\overrightarrow{PP_1} \times \overrightarrow{PP_2}$. Moreover, because the factor of one-half appears in both the numerator and denominator of equation 1, we can ignore this factor and calculate $A_0$ as

$$A_0 = |\overrightarrow{PP_1} \times \overrightarrow{PP_2}| \qquad (2a)$$
$$= \begin{vmatrix} x_1 - x & y_1 - y \\ x_2 - x & y_2 - y \end{vmatrix}$$
$$= (x_1 - x)(y_2 - y) - (x_2 - x)(y_1 - y)$$
$$= (y_1 - y_2)x + (x_2 - x_1)y + (x_1 y_2 - x_2 y_1)$$
$$= \alpha_0 x + \beta_0 y + \gamma_0$$

Values of $\alpha$, $\beta$ and $\gamma$ are analogously defined for computing $A_1$ and $A_2$. Equations for these triangle areas may be generated via cyclic permutation of the indices 0, 1 and 2

$$A_1 = |\overrightarrow{PP_2} \times \overrightarrow{PP_0}| = (y_2 - y_0)x + (x_0 - x_2)y + (x_2 y_0 - x_0 y_2) = \alpha_1 x + \beta_1 y + \gamma_1 \quad (2b)$$
$$A_2 = |\overrightarrow{PP_0} \times \overrightarrow{PP_1}| = (y_0 - y_1)x + (x_1 - x_0)y + (x_0 y_1 - x_1 y_0) = \alpha_2 x + \beta_2 y + \gamma_2$$

To calculate the barycentric coordinates for the pixel that corresponds to point $P$, the three areas $A_0$, $A_1$ and $A_2$ are calculated via equation 2 then substituted into equation 1. Interpolation is then performed using these barycentric coordinates as follows for the texture coordinates $(u, v)$

$$u = b_0 u_0 + b_1 u_1 + b_2 u_2$$
$$v = b_0 v_0 + b_1 v_1 + b_2 v_2 \qquad (3)$$

Interpolation of the color coordinates $(r, g, b)$ may be performed in a similar manner. However, interpolation that is performed in this way is logically inconsistent because the barycentric coordinates are defined in perspective space but the color and texture coordinates are defined in world space. The perspective transformation is non-linear, so the linear interpolation that is expressed by equation 3 produces errors. It is therefore necessary to undo the perspective transformation prior to interpolation in a manner [4] that is known as either rational linear interpolation [5] or hyperbolic interpolation [6] and that is shown as follows for the texture coordinate $u$

$$u = \frac{b_0(u_0/w_0) + b_1(u_1/w_1) + b_2(u_2/w_2)}{b_0(1/w_0) + b_1(1/w_1) + b_2(1/w_2)}$$

$$= \frac{b_0 w_1 w_2 u_0 + b_1 w_2 w_0 u_1 + b_2 w_0 w_1 u_2}{b_0 w_1 w_2 + b_1 w_2 w_0 + b_2 w_0 w_1} \quad (4)$$

The above equation suggests that equation 1 may be modified to express barycentric coordinates that are transformed from perspective space to eye space as follows

$$b_0' = w_1 w_2 A_0 / (w_1 w_2 A_0 + w_2 w_0 A_1 + w_0 w_1 A_2)$$
$$b_1' = w_2 w_0 A_1 / (w_1 w_2 A_0 + w_2 w_0 A_1 + w_0 w_1 A_2) \quad (5)$$
$$b_2' = w_0 w_1 A_2 / (w_1 w_2 A_0 + w_2 w_0 A_1 + w_0 w_1 A_2)$$

Because eye space and world space are related by a linear transformation, the transformed barycentric coordinates $b_0'$, $b_1'$ and $b_2'$ may be employed to perform linear interpolation of the texture coordinates $(u, v)$ or the color coordinates $(r, g, b)$ as shown as follows for the texture coordinates $(u, v)$

$$u = b_0' u_0 + b_1' u_1 + b_2' u_2$$
$$v = b_0' v_0 + b_1' v_1 + b_2' v_2 \quad (6)$$

Equations 2, 5 and 6 permit derivation of a set of equations that may be used for the "MIP mapping" [7] or pre-filtered approach to texture mapping that requires the partial derivatives [8] $\partial u/\partial x$, $\partial u/\partial y$, $\partial v/\partial x$ and $\partial v/\partial y$. Equation 3 implies that

$$u = f_u(b_0', b_1', b_2')$$
$$v = f_v(b_0', b_1', b_2') \qquad (7)$$

And equations 2 and 5 imply that

$$b_i' = g_i(x, y) \qquad (8)$$

Given equations 7 and 8, the chain rule of partial differentiation yields [9,10]

$$\frac{\partial u}{\partial x} = \frac{\partial f_u}{\partial b_0'}\frac{\partial b_0'}{\partial x} + \frac{\partial f_u}{\partial b_1'}\frac{\partial b_1'}{\partial x} + \frac{\partial f_u}{\partial b_2'}\frac{\partial b_2'}{\partial x}$$

$$= \frac{w_1 w_2 \alpha_0 (u_0 - u) + w_2 w_0 \alpha_1 (u_1 - u) + w_0 w_1 \alpha_2 (u_2 - u)}{w_1 w_2 A_0 + w_2 w_0 A_1 + w_0 w_1 A_2}$$

$$\frac{\partial u}{\partial y} = \frac{\partial f_u}{\partial b_0'}\frac{\partial b_0'}{\partial y} + \frac{\partial f_u}{\partial b_1'}\frac{\partial b_1'}{\partial y} + \frac{\partial f_u}{\partial b_2'}\frac{\partial b_2'}{\partial y}$$

$$= \frac{w_1 w_2 \beta_0 (u_0 - u) + w_2 w_0 \beta_1 (u_1 - u) + w_0 w_1 \beta_2 (u_2 - u)}{w_1 w_2 A_0 + w_2 w_0 A_1 + w_0 w_1 A_2}$$

$$\frac{\partial v}{\partial x} = \frac{\partial f_v}{\partial b_0'}\frac{\partial b_0'}{\partial x} + \frac{\partial f_v}{\partial b_1'}\frac{\partial b_1'}{\partial x} + \frac{\partial f_v}{\partial b_2'}\frac{\partial b_2'}{\partial x}$$

$$= \frac{w_1 w_2 \alpha_0 (v_0 - v) + w_2 w_0 \alpha_1 (v_1 - v) + w_0 w_1 \alpha_2 (v_2 - v)}{w_1 w_2 A_0 + w_2 w_0 A_1 + w_0 w_1 A_2}$$

$$\frac{\partial v}{\partial y} = \frac{\partial f_v}{\partial b_0'}\frac{\partial b_0'}{\partial y} + \frac{\partial f_v}{\partial b_1'}\frac{\partial b_1'}{\partial y} + \frac{\partial f_v}{\partial b_2'}\frac{\partial b_2'}{\partial y}$$

$$= \frac{w_1 w_2 \beta_0 (v_0 - v) + w_2 w_0 \beta_1 (v_1 - v) + w_0 w_1 \beta_2 (v_2 - v)}{w_1 w_2 A_0 + w_2 w_0 A_1 + w_0 w_1 A_2} \qquad (9)$$

A useful form of this equation may be derived by taking both $\partial u$ and $\partial v$ to be the pixel spacing $S$, then by multiplying both sides of the equation by $S$ to obtain the total differentials $du$ and $dv$ that represent the instantaneous rate of change of $u$ and $v$ at the pixel

$$du = S \frac{w_1 w_2 (\alpha_0 + \beta_0)(u_0 - u) + w_2 w_0 (\alpha_1 + \beta_1)(u_1 - u) + w_0 w_1 (\alpha_2 + \beta_2)(u_2 - u)}{w_1 w_2 A_0 + w_2 w_0 A_1 + w_0 w_1 A_2}$$

$$dv = S \frac{w_1 w_2 (\alpha_0 + \beta_0)(v_0 - v) + w_2 w_0 (\alpha_1 + \beta_1)(v_1 - v) + w_0 w_1 (\alpha_2 + \beta_2)(v_2 - v)}{w_1 w_2 A_0 + w_2 w_0 A_1 + w_0 w_1 A_2}$$

In this equation, $u$ and $v$ are the values interpolated at the pixel via equation 6, $\alpha$ and $\beta$ are defined by equation 2, and the denominator is defined by equation 5. Inspection of equations 5 and 6 reveals that the setup computation for each triangle requires creation of the nine terms $w_1 w_2 \alpha_0$, $w_1 w_2 \beta_0$, $w_1 w_2 \gamma_0$, $w_2 w_0 \alpha_1$, $w_2 w_0 \beta_1$, $w_2 w_0 \gamma_1$, $w_0 w_1 \alpha_2$, $w_0 w_1 \beta_2$ and $w_0 w_1 \gamma_2$. Following this setup computation, the areas of triangles $PP_1 P_2$, $PP_2 P_0$ and $PP_0 P_1$ may be computed on a per-pixel basis as

$$A_0 = w_1 w_2 \alpha_0 x + w_1 w_2 \beta_0 y + w_1 w_2 \gamma_0$$
$$A_1 = w_2 w_0 \alpha_1 x + w_2 w_0 \beta_1 y + w_2 w_0 \gamma_1$$
$$A_2 = w_0 w_1 \alpha_2 x + w_0 w_1 \beta_2 y + w_0 w_1 \gamma_2 \qquad (11)$$

The areas can then be used on a per-pixel basis to compute barycentric coordinates $b_0'$, $b_1'$ and $b_2'$ via equation 5, then those barycentric coordinates can be used to linearly interpolate color coordinates $(r, g, b)$ and texture coordinates $(u, v)$ as shown for the texture coordinates in equation 6, as well as to calculate the partial derivatives and total differentials via equations 9 and 10.

One further optimization is possible. It is necessary to compute the area equations of equation 11 for only the first pixel shaded for each triangle, and therefore this computation may be included in the setup computation for each triangle. Thereafter, these areas may be updated by addition or subtraction via a delta calculation. For example, if the next pixel is located one pixel (i.e., one unit) removed in the positive *x*-direction, then the areas may be updated by addition of $w_1 w_2 \alpha_0$, $w_2 w_0 \alpha_1$ and $w_0 w_1 \alpha_2$ to the results for the previous pixel. Similarly, if the next pixel is located one pixel away in the negative y-direction, then the areas may be updated by subtraction of $w_1 w_2 \beta_0$, $w_2 w_0 \beta_1$ and $w_0 w_1 \beta_2$ from the results for the previous pixel. So a scan conversion algorithm that proceeds from one pixel to an adjacent pixel permits calculation of barycentric coordinate interpolants via five additions, one reciprocal and three multiplies, as can be appreciated from inspection of equations 5 and 11.

The three products $w_1 w_2$, $w_2 w_0$ and $w_0 w_1$ appear in both the numerator and denominator of equations 5, 9 and 10, so it is possible to block normalize these products and keep only the mantissas of the normalization. For example, if the *w*-coordinates were defined as 15-bit integers, then the products would be 30-bit integers that could be block normalized to 15-bit integers, discarding the exponent of the normalization. Further block normalization may be possible for the nine terms $w_1 w_2 \alpha_0$, $w_1 w_2 \beta_0$, $w_1 w_2 \gamma_0$, $w_2 w_0 \alpha_1$, $w_2 w_0 \beta_1$, $w_2 w_0 \gamma_1$, $w_0 w_1 \alpha_2$, $w_0 w_1 \beta_2$ and $w_0 w_1 \gamma_2$. The aim of this block normalization is to permit the use of integer representation in place of floating-point representation.

**References**

1. Blinn JF. W pleasure, w fun. *IEEE Computer Graphics and Applications*, 18(3): 78-82, 1998.
2. Barnhill RE. Representation and approximation of surfaces. In Rice JR (ed.), *Mathematical Software III*, 69-120, Academic Press, New York, NY, 1977.
3. Kelly PJ, Straus EG. *Elements of Analytic Geometry*, 140, Scott, Forseman and Company, Glenview, IL, 1970.
4. Fuchs H, Goldfeather J, Hultquist JP, Spach S, Austin JD, Brooks II FP, Eyles JG, Poulton J. Fast spheres, shadow, textures, transparencies and image enchancement in pixel-planes. *Computer Graphics*, 19(3): 111-120, 1985.
5. Heckbert PS, Moreton HP. Interpolation for polygon texture mapping and shading. In Rogers DF, Earnshaw RA (eds.), *State of the Art in Computer Graphics: Visualization and Modeling*, 101-111, Springer-Verlag, New York NY, 1991.
6. Blinn JF. Hyperbolic interpolation. *IEEE Computer Graphics and Applications*, 12(4): 89-94, 1992.
7. Williams L. Pyramidal parametrics. *Computer Graphics* 17(3): 1-11, 1983.
8. Heckbert P. Texture mapping polygons in perspective. *NYIT Computer Graphics Lab Technical Memo #13*, 1983.
9. Thomas, GB Jr. The chain rule for partial derivatives. In: *Calculus and Analytic Geometry* 3rd ed., 683-687, Addison-Wesley, Reading, MA, 1960.
10. Watkins GS, Brown RA. System for polygon interpolation using instantaneous values in a variable. *United States Patent 5361386*, 1994.